\documentclass[seceq,preprint]{ptptex}

\usepackage{amsmath}
\usepackage{amssymb}

\usepackage{graphicx}
\usepackage{color}

\usepackage{float}

\preprintnumber{IU-TH-15}

\title{
Complex action suggests future-included theory
}

\author{%
Keiichi \textsc{Nagao}\footnote{E-mail: keiichi.nagao.phys@vc.ibaraki.ac.jp}
and Holger Bech \textsc{Nielsen}\footnote{E-mail: hbech@nbi.dk}
}

\inst{
$^{*)}$Faculty of Education, Ibaraki University, Bunkyo 2-1-1, Mito 310-8512 
Japan \\
{\it and }\\
$^{**)}$Niels Bohr Institute, University of Copenhagen, 
Blegdamsvej 17, Copenhagen $\O$, Denmark
}

\abst{

In quantum theory its action is usually taken to be real, but we can consider another theory 
whose action is complex. In addition, in the Feynman path integral, 
the time integration is usually performed over the period between the initial time $T_A$ and 
some specific time, say, the present time $t$. Besides such a future-not-included theory, 
we can consider the future-included theory, in which 
not only the past state $| A(T_A) \rangle$ at the initial time $T_A$ 
but also the future state $| B(T_B) \rangle$ at the final time $T_B$ is given at first, 
and the time integration is performed over the whole period from the past to the future. 
Thus quantum theory can be classified into four types, 
according to whether its action is real or not, and whether the future is included or not. 
We argue that, if a theory is described with a complex action,  
then such a theory 
is suggested to be the future-included theory, rather than the future-not-included theory. 
Otherwise persons living at different times would see different histories of the universe.}

\begin{document}

\maketitle

\section{Introduction}

Quantum theory is usually described by using the the Feynman path integral (FPI), 
where the time integration is performed over the period between the initial time $T_A$ and 
some specific time, say, the present time $t$. 
In addition to this future-not-included theory, we can consider  
another formulation, the future-included theory, in which 
not only the past state $| A(T_A) \rangle$ at the initial time $T_A$ 
but also the future state $| B(T_B) \rangle$ at the final time $T_B$ is given at first, 
and the time integration is performed over the whole period from the past to the future. 
In addition, in quantum theory its action is usually taken to be real. 
Let us call this the real action theory (RAT). 
We can consider another theory whose action is complex at the fundamental level. 
If we pursue a fundamental theory, it is better to require fewer conditions to be imposed on it at first. 
In this sense such a complex action theory (CAT) is preferable to the RAT, 
because the former has fewer conditions by at least one: there is no reality condition on the action. 
Thus quantum theory can be classified into four types, 
according to whether its action is real or not, and whether the future is included or not, 
as summarized in Table~\ref{tab:four_theories}.

\begin{table}
\caption{Four types of quantum theory.}
\label{tab:four_theories}
\begin{center}
\begin{tabular}{|l|c|c|}
\hline
    & Real action & Complex action \\
\hline 
Future is not included.  &  Future-not-included RAT  &  Future-not-included CAT \\  
\hline  
Future is included. &  Future-included RAT  &  Future-included CAT  \\   
\hline 
\end{tabular}
\end{center}
\end{table}
%

We have studied various properties of both the future-included and future-not-included CAT. 
In particular, the future-included CAT has been investigated 
with the expectation that the imaginary part of the action 
would give some falsifiable 
predictions~\cite{Bled2006, Nielsen:2005ub, Nielsen:2007ak, Nielsen:2008cm},  
and various interesting suggestions have been made for the Higgs mass~\cite{Nielsen:2007mj}, 
quantum-mechanical philosophy~\cite{newer1,Vaxjo2009,newer2}, 
some fine-tuning problems~\cite{Nielsen2010qq,degenerate}, 
black holes~\cite{Nielsen2009hq}, 
de Broglie--Bohm particles and a cut-off in loop diagrams~\cite{Bled2010B}. 
In addition, in Ref.~\cite{Nagao:2010xu}, introducing the proper inner product $I_Q$ 
for the Hamiltonian $\hat{H}$\footnote{
$\hat{H}$ is generically non-normal.  Hence the set of the Hamiltonians that we considered is much larger 
than that of the PT-symmetric non-Hermitian Hamiltonians, 
which has been intensively studied~\cite{Bender:1998ke,Bender:1998gh,Bender:2011ke,Mostafazadeh_CPT_ip_2002,
Mostafazadeh_CPT_ip_2003}. },  
where a Hermitian operator $Q$\footnote{
In the special case of the Hamiltonian $\hat{H}$ 
being normal, $Q$ is just a unit operator. }  is chosen 
so that the eigenstates of $\hat{H}$ 
become orthogonal to each other 
with respect to $I_Q$\footnote{Similar inner products are also studied in Refs.~\cite{Geyer, Mostafazadeh_CPT_ip_2002, Mostafazadeh_CPT_ip_2003}. }, 
we showed that we can effectively 
obtain a Hamiltonian that is $Q$-Hermitian, i.e., 
Hermitian with respect to $I_Q$,  after a long time development. 
Furthermore, using the complex coordinate 
formalism~\cite{Nagao:2011za}, 
we explicitly 
derived the momentum relation $p=m \dot{q}$, where $m$ is a complex mass, 
via the FPI~\cite{Nagao:2011is}.

In the future-included CAT, 
the normalized matrix element~\cite{Bled2006}\footnote{The normalized matrix element $\langle \hat{\cal O} \rangle^{BA}$ 
is called the weak value~\cite{AAV} in the context of the future-included RAT, 
and it has been intensively studied. 
For details of the weak value, see Refs.\cite{AAV, review_wv} and references therein.} 
$\langle \hat{\cal O} \rangle^{BA} 
\equiv \frac{ \langle B(t) |  \hat{\cal O}  | A(t) \rangle }{ \langle B(t) | A(t) \rangle }$, 
where $t$ is an arbitrary time ($T_A \leq t \leq T_B$), 
is a strong candidate for an expectation value 
of the operator $\hat{\cal O}$. 
Indeed, if we regard $\langle \hat{\cal O} \rangle^{BA}$ 
as the expectation value in the future-included CAT, 
we can obtain the Heisenberg equation, Ehrenfest's theorem, 
and a conserved probability current density~\cite{Nagao:2012mj,Nagao:2012ye}. 
Utilizing the mechanism for effectively obtaining a $Q$-Hermitian Hamiltonian~\cite{Nagao:2010xu}, 
we proposed the correspondence principle, 
which claims that, 
if we regard $\langle \hat{\cal O} \rangle^{BA}$ 
as an expectation value in the future-included CAT,  
the expectation value at the present time $t$ for large $T_B-t$ and large $t- T_A$ 
corresponds to 
that of the future-not-included theory with the proper inner product 
for large $t- T_A$~\cite{Nagao:2012mj,Nagao:2012ye}.  
Therefore, the future-included CAT, which influences the past in principle, 
is not excluded phenomenologically, though it looks very exotic.

As for the future-not-included CAT, 
an expectation value of an operator $\hat{\cal O}$ is given by 
$\langle \hat{\cal O} \rangle^{AA} 
\equiv \frac{ \langle A(t) |  \hat{\cal O}  | A(t) \rangle }{ \langle A(t) | A(t) \rangle }$. 
In Ref.~\cite{Nagao:2013eda}, 
we studied the various properties of 
$\langle O \rangle^{AA}$, 
and pointed out that the momentum relation $p=m \dot{q}$, which was shown to be correct 
in the future-included CAT~\cite{Nagao:2011is}, 
is not valid in the future-not-included CAT. 
Looking at the time development of 
$\langle O \rangle^{AA}$, 
we obtained the correct momentum relation in the future-not-included CAT, 
$p= \left( m_R + m_I^2 / m_R \right) \dot{q}$, 
where $m_R$ and $m_I$ are the real and imaginary parts of $m$ respectively. 
We also argued that its classical theory 
is described by a certain real action $S_{\text{eff}}$. 
In addition, we provided another way to understand the time development of 
the future-not-included theory by making use of the future-included theory. 
Furthermore, applying the method 
of deriving the momentum relation via the FPI~\cite{Nagao:2011is} 
to the future-not-included theory properly by introducing a formal Lagrangian, 
we derived the correct momentum relation in the future-not-included theory, 
which is consistent with that mentioned above.

Thus the future-not-included CAT has very intriguing properties, 
so it seems to be worthwhile to study it more. 
However, 
in this letter, we point out that, if we adopt a theory whose action is complex, 
then it is suggested that the theory has to be the future-included CAT, rather than 
the future-not-included CAT. 
We encounter a philosophical discrepancy in the future-not-included CAT. 
We illustrate this suggestion with a couple of simple examples after briefly 
reviewing the future-included and future-not-included CAT.

\section{Review of the future-included and future-not-included CAT}

In a system defined with a single degree of freedom, 
we consider the CAT, in which the FPI is described with the Lagrangian 
$L(q(t), \dot{q}(t))=\frac{1}{2}m \dot{q}^2- V(q)$, 
where $m$ is a complex mass, and  
$V(q)$ is a complex potential term.

Following Refs.\cite{Nagao:2012mj,Nagao:2012ye,Nagao:2017book}, 
we briefly review the future-included theory.  
In the future-included theory, not only the past state $| A(T_A) \rangle$ at the initial time $T_A$ 
but also the future state $| B(T_B) \rangle$ at the final time $T_B$ are given at first, 
and $| A(t) \rangle$ and $| B(t) \rangle$ are supposed to time-develop according to 
the Schr\"{o}dinger equations 
\begin{eqnarray}
&&i \hbar \frac{d}{d t} | A(t) \rangle = \hat{H} | A(t) \rangle , 
\label{schro_eq_Astate} \\
&&i \hbar \frac{d}{d t} | B(t) \rangle = \hat{H}^\dag | B(t) \rangle . 
\label{schro_eq_Bstate} 
\end{eqnarray}
In Refs.\cite{Nagao:2012mj,Nagao:2012ye} 
we investigated the normalized matrix element 
$\langle \hat{\cal O} \rangle^{BA} \equiv 
\frac{ \langle B(t) |  \hat{\cal O}  | A(t) \rangle }{ \langle B(t) | A(t) \rangle }$\cite{Bled2006}, 
which is a strong candidate for an expectation value in the future-included theory. 
Indeed, this $\langle \hat{\cal O} \rangle^{BA}$ obeys 
$\frac{d}{dt} \langle \hat{\cal O} \rangle^{BA} 
= \langle  \frac{i}{\hbar} [ \hat{H} , \hat{\cal O} ]   \rangle^{BA}$. 
Substituting $\hat{q}_\text{new}$ and 
$\hat{p}_\text{new}$\footnote{$\hat{q}_\text{new}$ and $\hat{p}_\text{new}$ are  
generalized coordinate and momentum operators that are constructed 
in the context of the complex coordinate formalism~\cite{Nagao:2011za,Nagao:2017book} 
so that they are non-Hermitian and have complex eigenvalues $q$ and $p$. 
The complex coordinate formalism is not relevant for the purposes of this letter, 
so we do not discuss it. The details are referred to in Refs.~\cite{Nagao:2011za,Nagao:2017book}.} 
for $\hat{\cal O}$, 
we obtain 
\begin{eqnarray}
&&\frac{d}{dt} \langle \hat{q}_\text{new} \rangle^{BA} 
= \frac{1}{m} \langle \hat{p}_\text{new} \rangle^{BA} ,  \label{dqcdt}  \\
&&\frac{d}{dt} \langle \hat{p}_\text{new} \rangle^{BA} 
= - \langle V'(\hat{q}_\text{new}) \rangle^{BA} ,    \label{dpcdt} 
\end{eqnarray}
and Ehrenfest's theorem, 
$m\frac{d^2}{dt^2} \langle \hat{q}_\text{new} \rangle^{BA} 
= - \langle V'(\hat{q}_\text{new}) \rangle^{BA}$. 
Also, Eq.(\ref{dqcdt}) leads to the momentum relation 
$p=\frac{\partial L}{\partial \dot{q}} = m \dot{q}$. 
Thus, $\langle \hat{\cal O} \rangle^{BA}$ provides 
the simple time development of the saddle point for $\exp(\frac{i}{\hbar} S)$. 
In addition, using both the complex coordinate formalism\cite{Nagao:2011za} 
and the automatic hermiticity mechanism\cite{Nagao:2010xu,Nagao:2011za}, 
i.e., the mechanism to obtain a Hermitian Hamiltonian 
after a long time development, 
we obtained a correspondence principle that 
$\langle \hat{\cal O} \rangle^{BA}$ for large $T_B-t$ and large $t-T_A$ is almost 
equivalent to 
$\langle \hat{\cal O} \rangle_{Q'}^{AA}\equiv \frac{  \langle A(t) |_{Q'} \hat{\cal O}  | A(t) \rangle }
{ \langle A(t) |_{Q'} A(t) \rangle }$ for large $t-T_A$, 
where $Q'$ is a Hermitian operator that is used to define the proper 
inner product so that the eigenstates of the Hamiltonian become orthogonal to each other 
with regard to it. 
Thus the future-included theory is not excluded phenomenologically, though it looks 
very exotic.

Following Refs.\cite{Nagao:2010xu,Nagao:2011za,Nagao:2013eda,Nagao:2017book}, 
we briefly review the future-not-included theory.  
In the future-not-included theory, 
only the past state $| A(T_A) \rangle$ at the initial time $T_A$ 
is given at first, 
and $| A(t) \rangle$ is supposed to time-develop according to Eq.(\ref{schro_eq_Astate}).  
The expectation value in the future-not-included theory is given by  
$\langle \hat{\cal O} \rangle^{AA} 
\equiv \frac{  \langle A(t) | \hat{\cal O}  | A(t) \rangle }
{ \langle A(t) | A(t) \rangle } 
= {}_{N} \langle A(t) | \hat{\cal O} | A(t) \rangle_{N}$, 
where we have introduced a normalized state 
$|A(t) \rangle_{N} \equiv \frac{1}{\sqrt{ \langle {A}(t) | ~{A}(t) \rangle} } | {A}(t) \rangle$. 
Then, $| A(t) \rangle_{N}$ obeys the slightly modified Schr\"{o}dinger equation,
\begin{eqnarray}
i\hbar  \frac{d}{d t} | A(t) \rangle_{N} 
&=& \hat{H} | A(t) \rangle_{N} 
-{}_{ N} \langle A(t) | \hat{H}_{a} | A(t) \rangle_{N} | A(t) \rangle_{N} \nonumber \\
&=& \hat{H}_{h} | A(t) \rangle_{N} 
+ \left( \hat{H}_{a} -{}_{N} \langle A(t) | \hat{H}_{a} | A(t) \rangle_{N} \right) 
| A(t) \rangle_{N} , \label{sch_fni}
\end{eqnarray} 
where $\hat{H}_{h}$ and $\hat{H}_{a}$ are the Hermitian and anti-Hermitian parts of 
$\hat{H}$ respectively. 
In Eq.(\ref{sch_fni}) we see that 
the effect of the anti-Hermitian part of $\hat{H}$ disappears in the classical limit, 
though the theory is defined with $\hat{H}$ at the quantum level. 
In addition, we find the time development of $\langle \hat{\cal O}  \rangle^{A A}$ as follows: 
\begin{eqnarray}
i \hbar \frac{d}{d t} \langle \hat{\cal O}  \rangle^{A A} 
&=&
\langle [\hat{\cal O}, \hat{H}_h] \rangle^{A A} + \langle F(\hat{\cal O}, \hat{H}_a) 
\rangle^{A A}  
\simeq
\langle [\hat{\cal O}, \hat{H}_h] \rangle^{A(t) A(t)} ,  \label{ihbardeldeltOAA_t}
\end{eqnarray}
where $F(\hat{\cal O}, \hat{H}_a)(t)$, a quantum fluctuation term given by 
$F(\hat{\cal O}, \hat{H}_a)(t) 
= \left\{ \hat{\cal O} , \hat{H}_a  - \langle \hat{H}_a \rangle^{A A} \right\}  
= \left\{ \hat{\cal O} -\langle\hat{\cal O} \rangle^{A A} , \hat{H}_a  \right\}$, 
disappears in the classical limit. 
Substituting $\hat{q}_\text{new}$ and $\hat{p}_\text{new}$ 
for $\hat{\cal O}$ in Eq.(\ref{ihbardeldeltOAA_t}), we obtain 
\begin{eqnarray} 
\frac{d}{d t} \langle \hat{q}_\text{new}  \rangle^{A A} 
&\simeq& \frac{1}{ i \hbar   }
\langle [\hat{q}_\text{new} , \hat{H}_h] \rangle^{A A}  
\simeq \frac{1}{m_{\text{eff}}} 
\langle \hat{p}_\text{new} \rangle^{A A}  ,  \label{ihbardeldeltqhatAA_t} \\ 
\frac{d}{dt} \langle \hat{p}_\text{new}  \rangle^{A A} 
&\simeq&
\langle [\hat{p}_\text{new} , \hat{H}_h] \rangle^{A A}  
\simeq
- \langle V_R'( \hat{q}_\text{new} ) \rangle^{A A}  ,   \label{ihbardeldeltphatAA_t}
\end{eqnarray}
where 
$m_{\text{eff}} \equiv m_R + \frac{m_I^2}{m_R}$, 
and $V_R$ is the real part of the potential term $V$. 
Combining Eq.(\ref{ihbardeldeltqhatAA_t}) with Eq.(\ref{ihbardeldeltphatAA_t}), 
we obtain Ehrenfest's theorem, 
$m_{\text{eff}} \frac{d^2}{d t^2} \langle \hat{q}_\text{new}  \rangle^{A A}  
\simeq - \langle V_R'( \hat{q}_\text{new} ) \rangle^{A A}$, 
which suggests that the classical theory of the future-not-included theory 
is described not by the full action $S$, but 
$S_{\text{eff}}\equiv\int_{T_A}^t dt L_{\text{eff}}$, 
where $L_{\text{eff}}(\dot{q}, q) 
\equiv \frac{1}{2} m_{\text{eff}} \dot{q}^2 - V_{R}(q)$. 
Thus the classical theory of the future-not-included theory 
is described by $\delta S_{\text{eff}} = 0$. 
We also find that 
Eq.(\ref{ihbardeldeltqhatAA_t}) leads to the momentum relation 
$p= \frac{\partial L_{\text{eff}}}{\partial \dot{q}}= m_{\text{eff}} \dot{q}$.

We give a brief summary of the future-included and future-not-included theories 
in Table~\ref{tab:comparison_fi_fni}\cite{Nagao:2013eda}. 
We see that the classical theory of the future-included theory is quite in contrast to 
that of the future-not-included theory.

\begin{table}
\caption[AAA]{Comparison between the future-included and future-not-included theories.}
\label{tab:comparison_fi_fni}
\begin{center}
\begin{tabular}{|p{3.5cm}|p{4cm}|p{7cm}|}
\hline
    & future-included theory & future-not-included theory \\
\hline
action  &  $S=\int_{T_A}^{T_B}dt L$   & $S=\int_{T_A}^{t} dt L$ \\  
\hline
``expectation value" & $\langle \hat{\cal O} \rangle^{BA} = 
\frac{ \langle B(t) |  \hat{\cal O}  | A(t) \rangle }{ \langle B(t) | A(t) \rangle }$ & $\langle \hat{\cal O} \rangle^{AA} 
= \frac{  \langle A(t) | \hat{\cal O}  | A(t) \rangle }
{ \langle A(t) | A(t) \rangle }$  \\   
\hline
time development 
& $i \hbar \frac{d}{d t} \langle \hat{\cal O} \rangle^{BA}$  
& $i \hbar \frac{d}{d t} \langle \hat{\cal O}  \rangle^{AA}$ \\
& $= \langle  [ \hat{\cal O} , \hat{H} ]\rangle^{BA}$
& 
$= \langle [\hat{\cal O}, \hat{H}_h] \rangle^{AA} 
+ \langle \left\{ \hat{\cal O} -\langle\hat{\cal O} \rangle^{A A} , \hat{H}_a  \right\} \rangle^{A A}$  
\\ 
&
&
$\simeq \langle [\hat{\cal O}, \hat{H}_h] \rangle^{AA} $ \\
\hline
classical theory & $\delta S=0$ &  $\delta S_{\text{eff}}=0$,  $S_{\text{eff}}=\int_{T_A}^{t} dt L_{\text{eff}}$\\  
\hline
momentum relation  &  $p= m \dot{q}$   & $p= m_{\text{eff}} \dot{q}$ \\    
\hline
\end{tabular}
\end{center}
\end{table}
%
%


\section{Complex action suggests a future-included theory}

In the FPI 
$\int {\cal D} \text{path}~ \psi_B^* \psi_A e^{ \frac{i}{\hbar} S[\text{path}] }$,  
the integrand that includes the action $S$ is expressed as 
$e^{ \frac{i}{\hbar} S[\text{path}] } 
= e^{ \frac{i}{\hbar} S_R[\text{path}] } e^{ -\frac{1}{\hbar} S_I[\text{path}] }$,  
where $S_R$ and $S_I$ are real and imaginary parts of $S$, respectively. 
Since $e^{ -\frac{1}{\hbar} S_I[\text{path}] }$ can have higher orders of magnitude than 
$e^{ \frac{i}{\hbar} S_R[\text{path}] }$ 
and the boundary wave functions $\psi_B^* \psi_A$, 
it is $S_I[\text{path}]$ that has the greatest influence on the selection of paths. 
Paths realizing smaller $S_I[\text{path}]$ 
are essentially favored and chosen.\footnote{In other words, paths with larger imaginary parts of the eigenvalues of the Hamiltonian $\hat{H}$ are favored and chosen.} 
$S_I[\text{path}]$ is obtained by the time integration of the imaginary part of the Lagrangian 
$L_I(q, \dot{q}) \equiv \frac{1}{2}m_I \dot{q}^2 - V_I(q)$, where $m_I$ and $V_I$ 
are imaginary parts of $m$ and $V$, respectively. 
We symbolically write $L_I(q, \dot{q})$ as the function of $t$, $L_I^\text{path}(t)$. 
$S_I([T_A, t]) \equiv \int_{T_A}^t L_I^\text{path}(t') dt'$  
and 
$S_I([T_A, T_B]) = \int_{T_A}^{T_B} L_I^\text{path}(t') dt'$ 
are used in the future-not-included and future-included theories, respectively. 
If $L_I(q, \dot{q})$ varies greatly in time\footnote{A time-dependent non-Hermitian Hamiltonian is studied 
in Ref.~\cite{Fukuma:2013mx}.}, 
paths are nontrivially chosen in the FPI. 
We give a couple of simple examples of two paths, and discuss which path is chosen 
by comparing $S_I[\text{path}]$ in each example.

In the following, taking the initial time $T_A$ as $T_A=0$ for simplicity, 
we consider a pair of constant $L_I$ as the first example of two paths 
for pedagogical reasons. 
Next we present the second example, 
where one of the two $L_I$ is constant, but the other is time-dependent.  
In this second example, we show that, if we stand on the future-not-included theory and respect objectivity, 
then we encounter a philosophical contradiction, 
and thus we are led to the future-included theory.

\begin{figure}
 \centering
\includegraphics[width=10cm,bb=0 0 534.000000 312.600000]{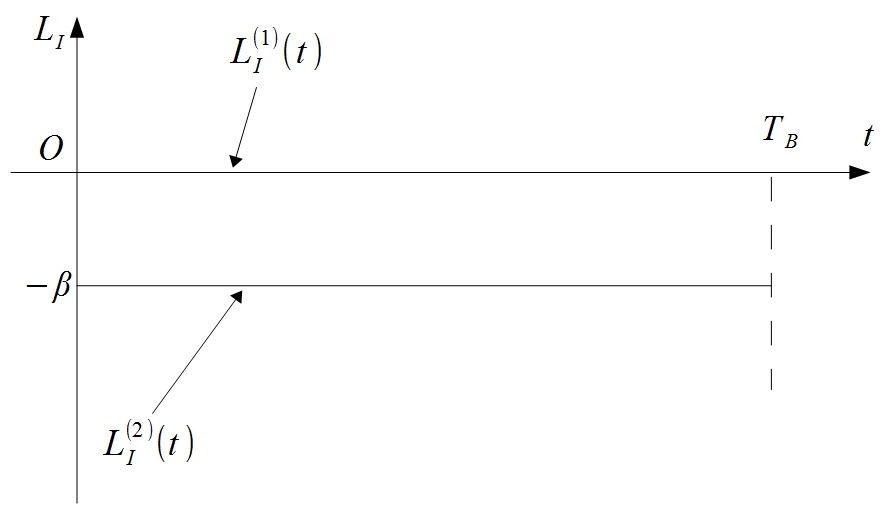} 
 \caption{The first example of $L_I$.}
 \label{fig:simplest_example_L_I}
\end{figure}

Let us begin with the first example,  
a pair of constant $L_I$ as two paths. 
Such a pair of  $L_I$ is defined as follows: 
$L_I^{(1)}(t) = 0$, 
$L_I^{(2)}(t) = -\beta$, 
where $\beta>0$. 
$L_I^{(1)}$ and $L_I^{(2)}$ 
are drawn in Fig.~\ref{fig:simplest_example_L_I}. 
Each $S_I^{(j)}$ for $L_I^{(j)}$ $(j=1,2)$ 
in the future-not-included theory is given by 
$S_I^{(1)}([0, t]) = \int_0^t L_I^{(1)}(t') dt' = 0$ and 
$S_I^{(2)}([0, t]) = \int_0^t L_I^{(2)}(t') dt'  = -\beta t$. 
Since $S_I^{(2)}  ([0, t]) < S_I^{(1)} ([0, t])$, 
a person living in the time $t$ who believes that our universe is described 
by the future-not-included theory judges that path $2$ is favored, 
and thinks that our universe is determined by path $2$. 
If another person believes the future-included theory, he compares 
$S_I^{(1)} ([0, T_B])  =0$ and  
$S_I^{(2)} ([0, T_B])  = -\beta T_B$. 
Since $S_I^{(2)}  ([0, T_B]) < S_I^{(1)} ([0, T_B])$, he judges that 
path $2$ is favored, and thinks that our universe is determined by path 2.  
This is a very simple example, so we do not encounter any problems.
Both interpretations, the future-included and future-not-included theories, can 
stand. 
However, if we consider a slightly more nontrivial example, then 
we could easily encounter difficulties. 
We see this in the next example.

\begin{figure}[h]
 \centering
\includegraphics[width=10cm,bb=0 0 531.600000 478.200000]{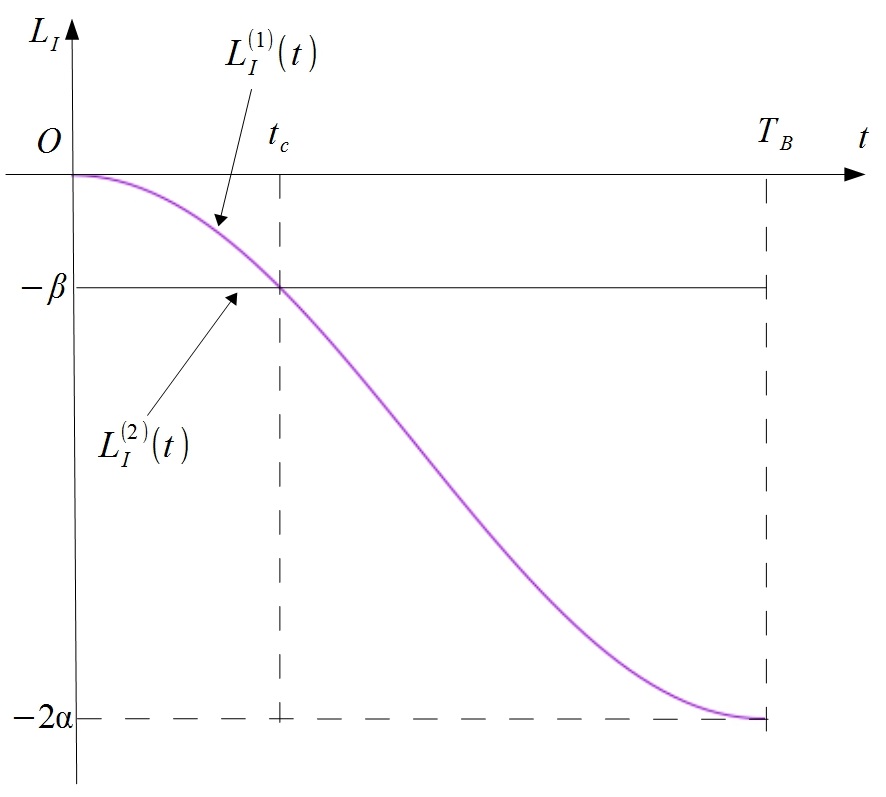}
 \caption{The second example of $L_I$.}
 \label{fig:example_L_I}
\end{figure}

Let us consider the second example 
such that one of $L_I$ varies in time. 
We take the following pair of $L_I$ as two paths: 
\begin{eqnarray}
&&L_I^{(1)}(t) =  \alpha \left\{ \cos \left(\frac{\pi}{T_B} t \right) -1 \right\} , \\
&&L_I^{(2)}(t) = -\beta , 
\end{eqnarray}
where $\alpha$ and $\beta$ are constants such that $\alpha > \beta>0$. 
$L_I^{(1)}$ and $L_I^{(2)}$ are drawn in Fig.~\ref{fig:example_L_I}, 
where $t_c$ is the solution to $L_I^{(1)}(t_c) = L_I^{(2)}(t_c)$, and found to be 
$t_c = \frac{T_B}{\pi} \cos^{-1}\left(1- \frac{\beta}{\alpha} \right)$. 
Let us suppose that a person living in the time $t$ believes the future-not-included theory. 
Each $S_I^{(j)}$ for $L_I^{(j)}$ $(j=1,2)$ is expressed as 
$S_I^{(1)} ([0, t]) 
= \alpha \left\{ \frac{T_B}{\pi} \sin \left(\frac{\pi}{T_B} t \right) - t \right\}$, 
$S_I^{(2)}  ([0, t]) = -\beta t$.

\begin{figure}[h]
 \centering
\includegraphics[width=10cm,bb=0 0 530.400000 481.200000]{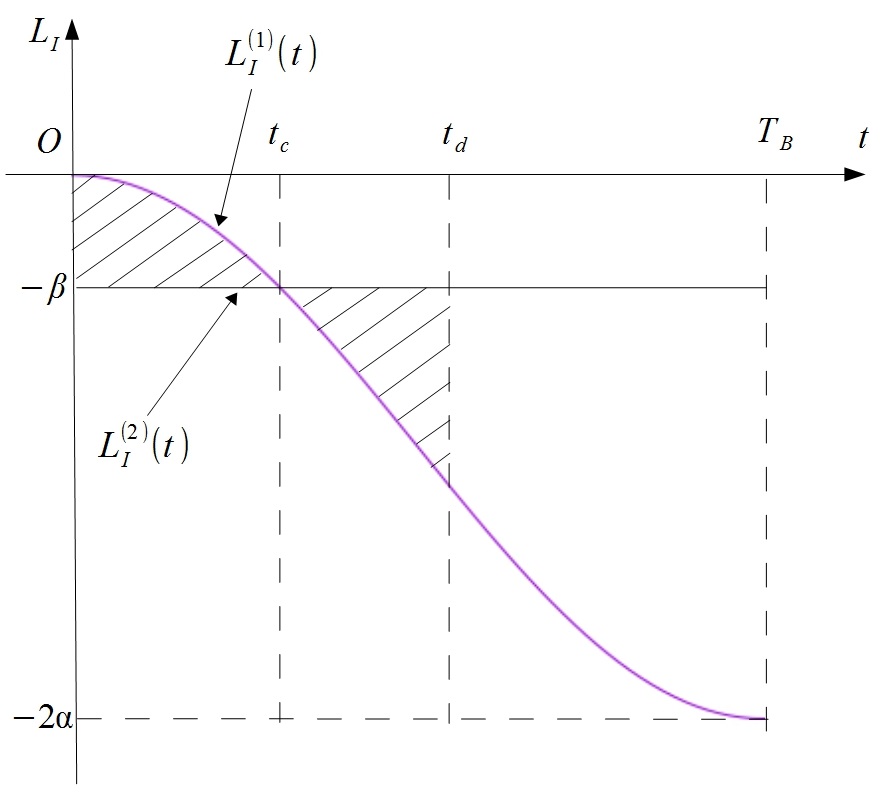} 
 \caption{The second example of $L_I$ with $t_d$ included.}
 \label{fig:example_L_I_w_td}
\end{figure}

At a glance, for $t<t_c$, 
we easily see that $S_I^{(2)}  ([0, t]) < S_I^{(1)} ([0, t])$, because $L_I^{(2)}(t) < L_I^{(1)}(t)$. 
So, for $t<t_c$, he judges that path $2$ is favored. 
Then how does he judge for $t>t_c$? 
We can answer this question by knowing the time $t_d$ 
such that $S_I^{(1)} ([0, t_d])$ balances with $S_I^{(2)}  ([0, t_d])$. 
That is, $t_d$ is defined as the solution to 
$S_I^{(1)} ([0, t_d]) = S_I^{(2)} ([0, t_d])$, 
which is reduced to 
$\sin \left(\frac{\pi}{T_B} t_d \right) = \left( 1 - \frac{\beta}{\alpha} \right) \frac{\pi}{T_B}t_d$. 
In Fig.~\ref{fig:example_L_I_w_td}, 
$t_d$ is 
determined so that each area of the two domains with slanted lines is equal to each other. 
Using this $t_d$, we find the following relations: 
\begin{eqnarray}
&&S_I^{(1)} ([0, t]) > S_I^{(2)} ([0, t])  \quad \text{for}~ 0 \le t< t_d , \label{SI1>SI2_t<td} \\
&&S_I^{(1)} ([0, t]) < S_I^{(2)} ([0, t]) \quad \text{for}~  t_d < t \le T_B. \label{SI1<SI2_t>td}
\end{eqnarray}
In the future-not-included theory only what happened in the past can matter. 
Therefore, the person living at the earlier time $0 \le t <t_d$ judges that path $2$ 
is chosen, but in the later time $t_d < t \le T_B$ he will judge that path $1$ is chosen. 
Thus we have encountered a strange situation. 
We usually want to have objectivity for any theory to be reasonable, but 
the mentioned property 
indicates that the future-not-included theory is subjective. 
Such a scenario in which what happened should depend on whom you ask, 
which lacks objectivity, 
reminds us of the so-called 
Mandela effect\footnote{That is, a large part of the 
population believed that deceased former South African President Nelson Mandela 
had already been dead a couple of decades before he really died~\cite{Mandela_effect}.}, 
which was named by the blogger Fiona Broome. 
If in the later time path $1$ is chosen, then even in the earlier time path $1$ 
should have been chosen, as long as we respect objectivity. 
Looking at the history, we will effectively find influence from the future 
looking back even in the future-not-included theory.  
This is a philosophical contradiction. 
To avoid this discrepancy, the person is led to  
the future-included theory, rather than the future-not-included theory.

Indeed, if he believes the future-included theory, then he compares 
$S_I^{(1)} ([0, T_B]) = -\alpha T_B$ and 
$S_I^{(2)}  ([0, T_B]) = -\beta T_B$. 
Since $S_I^{(1)}  ([0, T_B]) < S_I^{(2)} ([0, T_B])$, 
he judges that path $1$ is favored at any time $t$ ($0 \le t \le T_B$). 
We do not encounter any contradiction in the future-included theory. 
Therefore, if an action is allowed to be complex, 
then such an action has to be described in the future-included theory. 
It is very interesting that complex action suggests the future-included theory.

If the person persists in believing the future-not-included theory, how does he feel 
in the earlier time $0 \le t <t_d$? 
In the earlier time 
$0 \le t <t_d$, since $S_I^{(2)} ([0, t]) < S_I^{(1)} ([0, t])$, 
he thinks that it is a miraculous phenomenon that path $1$ is chosen. 
This story implies that, if the action of our universe is allowed to be complex, 
then we could see miraculous phenomena. 
Oppositely, if we see miraculous phenomena in the usual 
theory, i.e., the future-not-included RAT, then we have a possibility that our universe is 
described by the future-included CAT. 
If so, such phenomena can be understood reasonably well. 
The future-included CAT gives similar effects to the anthropic principle.

\section{Discussion}

In this letter, 
after briefly reviewing the future-included and future-not-included CAT, 
we have given a couple of examples of imaginary parts of Lagrangians $L_I$ as two paths, 
and discussed which path is favored and chosen by comparing 
imaginary parts of actions $S_I$. 
In one of the examples we have encountered a philosophical contradiction 
in the future-not-included CAT as long as we respect objectivity. 
In the future-not-included theory, as future becomes past, 
the influence of $L_I$ 
in such time intervals becomes relevant for the relative probability 
for various states in the FPI. 
This would lead to a strange re-choosing of initial states 
in the perspective of determinism so as to 
have had the smallest $S_I$ until the present time. 
Such changing of initial states would be exceedingly strange at least classically.
Indeed, in Ref.~\cite{Nagao:2013eda}, 
we reported such a complicated aspect 
of the future-not-included theory. 
We showed that time derivatives of 
$\langle \hat{q}_\text{new} \rangle_{AA}$ and $\langle \hat{p}_\text{new} \rangle_{AA}$ 
have complicated anticommutation terms, 
and provided an unusual way to understand the time development 
by using such re-choosing of the initial states. 
If a historian sees that people in the past were governed by their future, 
then it would be strange if we were not governed by the future. 
If we are to be governed by the future, then the future should exist. 
The historical people would have the happening leading to low $L_I$ in their 
future because $e^{ -\frac{1}{\hbar} S_I[\text{path}] }$ promotes it so. 
This means that they are influenced by the future. 
Thus we are led to the future-included CAT. 
If we stand on the future-included CAT, 
we do not see any contradiction. 
It is much stabler for the 
predictions and consistent with determinism to have influence 
from an always or ever-existing future. 
Therefore, if an action is allowed to be complex, then 
such an action has to be described in the future-included theory. 
Agreeing with determinism, at least crudely, is a major 
benefit of the future-included CAT. Also, the future-included CAT can yield a simpler classical 
equation of motion for $\langle \hat{q}_\text{new} \rangle_{BA}$ and 
$\langle \hat{p}_\text{new} \rangle_{BA}$ than the future-not-included CAT.

In the future-included theories we need a final condition analogous to an initial condition 
to deliver the final state $|B(T_B) \rangle$. 
In the future-included RAT we need two boundary conditions $|B(T_B) \rangle$ and 
$|A(T_A) \rangle$. So the future-included RAT is a bit more complicated than 
the future-not-included RAT that needs only one boundary condition. 
In the future-included CAT we obtain the boundaries unified with the dynamics; 
both $|B(T_B) \rangle$ and $|A(T_A) \rangle$ are effectively obtained from $S_I$. 
The future-included CAT makes such an initial or final condition automatically. 
Indeed, in Refs.~\cite{Nagao:2015bya, Nagao:2017cpl, Nagao:2017book, Nagao:2017ztx}, 
introducing a slightly modified normalized matrix element 
$\langle \hat{\cal O} \rangle_Q^{BA} 
\equiv \frac{ \langle B(t) |_Q  \hat{\cal O}  | A(t) \rangle }{ \langle B(t) |_Q A(t) \rangle }$, 
which 
is obtained just by changing the notation of $\langle B(t)|$ 
as $\langle B(t)| \rightarrow \langle B(t)|_Q \equiv \langle B(t)|Q$ 
in $\langle \hat{\cal O} \rangle^{BA}$, 
we presented a theorem that states that, 
provided that an operator $\hat{\cal O}$ is $Q$-Hermitian, 
the normalized matrix element 
$\langle \hat{\cal O} \rangle_Q^{BA}$ becomes real and 
time-develops under a $Q$-Hermitian Hamiltonian for  
$| B (t) \rangle$ and $| A (t) \rangle$ selected such 
that the absolute value of the transition amplitude  
$|\langle B(t) |_Q A(t) \rangle|$ 
is maximized. 
We call this way of thinking the maximization principle. 
This provides us both reality of $\langle \hat{\cal O} \rangle_Q^{BA}$ 
and $Q$-hermiticity of the Hamiltonian, 
even though $\langle \hat{\cal O} \rangle_Q^{BA}$ 
is generically complex by definition and the given Hamiltonian $\hat{H}$ is non-normal 
at first\footnote{In the RAT case, only reality of $\langle \hat{\cal O} \rangle^{BA}$ is the point, 
because the given $\hat{H}$ is Hermitian. }.  
We found that in the case of the CAT a unique class of paths is chosen by the maximization principle. 
Besides this fact, since the functional integral expression is simpler in the future-included theories than the future-not-included theories, 
we argued that the future-included CAT is the most elegant. 
The study in this letter partly supports this speculation.

In this letter we have argued that the existence of an imaginary part of the action suggests 
the future-included theory. 
Then, can we say the reverse, i.e., does the future-included theory suggest the existence of 
an imaginary part of the action? 
It is not clear, but it would be interesting if we could say something about it. 
If we show that the effects of the imaginary 
part turn out to be unobservable in practice in a good approximation, 
then we can argue that there is no strong reason to assume the action to be real in nature. 
The reality of the action can be regarded 
as a restriction on parameters in the action, 
and thus really an extra -- and according to our argument -- 
unnecessary assumption. 
So the real benefit from our CAT would be that we can have 
a more general action by getting rid of the restriction.


\section*{Acknowledgements}

K.N. would like to thank the members and visitors of NBI 
for their kind hospitality during his visits to Copenhagen. 
H.B.N. is grateful to NBI for allowing him to work there as emeritus. 
In addition, we acknowledge the TV personality Sidney Lee 
for having drawn our attention to the Mandela effect stories.




\begin{thebibliography}{9}



\bibitem{Bled2006}
H.~B.~Nielsen and M.~Ninomiya, 
Proc. Bled 2006: What Comes Beyond the Standard Models, 
pp.~87-124 (2006) 
[arXiv:hep-ph/0612250]. 









\bibitem{Nielsen:2007ak}
  H.~B.~Nielsen and M.~Ninomiya,
  Int.\ J.\ Mod.\ Phys.\  A {\bf 23}, 919 (2008). 
%


\bibitem{Nielsen:2008cm}
  H.~B.~Nielsen and M.~Ninomiya, 
  Int.\ J.\ Mod.\ Phys.\  A {\bf 24}, 3945 (2009).
%







\bibitem{Nielsen:2005ub}
  H.~B.~Nielsen and M.~Ninomiya,
  Prog.\ Theor.\ Phys.\  {\bf 116}, 851 (2006). 



\bibitem{Nielsen:2007mj}
H.~B.~Nielsen and M.~Ninomiya, 
Proc. Bled 2007: What Comes Beyond the Standard Models, pp.~144-85 (2007) [arXiv:0711.3080 [hep-ph]].



\bibitem{newer1}
  H.~B.~Nielsen and M.~Ninomiya,
  arXiv:0910.0359 [hep-ph]. 



\bibitem{Vaxjo2009}
H.~B.~Nielsen, 
Found. Phys. {\bf 41}, 608 (2011) [arXiv:0911.4005[quant-ph]]. 






\bibitem{newer2}
H.~B.~Nielsen and M.~Ninomiya, 
Proc. Bled 2010: What Comes Beyond the Standard Models, 
pp.~138-57 (2010) [arXiv:1008.0464 [physics.gen-ph]]. 

  





\bibitem{Nielsen2010qq}
H.~B.~Nielsen,
arXiv:1006.2455 [physic.gen-ph].


\bibitem{degenerate}
H.~B.~Nielsen and M.~Ninomiya,
arXiv:hep-th/0701018.





\bibitem{Nielsen2009hq}
  H.~B.~Nielsen,
arXiv:0911.3859 [gr-qc].




\bibitem{Bled2010B}
H.~B.~Nielsen, M.~S.~Mankoc~Borstnik, K.~Nagao, and G.~Moultaka, 
%
Proc. Bled 2010: What Comes Beyond the Standard Models, 
pp.~211-6 (2010) [arXiv:1012.0224 [hep-ph]]. 






\bibitem{Nagao:2010xu}
K.~Nagao and H.~B.~Nielsen,
Prog.\ Theor.\ Phys. {\bf 125}, 633 (2011).






\bibitem{Bender:1998ke}
  C.~M.~Bender and S.~Boettcher,
  Phys.\ Rev.\ Lett.\  {\bf 80}, 5243 (1998).

\bibitem{Bender:1998gh}
  C.~M.~Bender, S.~Boettcher, and P.~Meisinger,
  J.\ Math.\ Phys.\  {\bf 40}, 2201 (1999).


\bibitem{Bender:2011ke} 
  C.~M.~Bender and P.~D.~Mannheim,
  Phys.\ Rev.\ D {\bf 84}, 105038 (2011).





\bibitem{Mostafazadeh_CPT_ip_2002}
A. Mostafazadeh, 
J.\ Math.\ Phys.\ {\bf 43}, 3944 (2002).  
%



\bibitem{Mostafazadeh_CPT_ip_2003}
A. Mostafazadeh, 
J.\ Math.\ Phys.\ {\bf 44}, 974 (2003).





\bibitem{Geyer}
F. G. Scholtz, H. B. Geyer, and F. J. W. Hahne, Ann. Phys. {\bf 213}, 74 (1992). 







\bibitem{Nagao:2011za}
  K.~Nagao and H.~B.~Nielsen,
Prog.\ Theor.\ Phys. {\bf 126}, 1021 (2011); 
{\bf 127}, 1131 (2012) [erratum]. 





\bibitem{Nagao:2011is}
  K.~Nagao and H.~B.~Nielsen, 
   Int.\ J.\ Mod.\ Phys.\ A{\bf 27}, 1250076 (2012). 






		
\bibitem{AAV}
	Y. Aharonov, D. Z. Albert, and L. Vaidman,
	Phys. Rev. Lett. 
	{\bf 60}, 1351 (1988).





\bibitem{review_wv}
	Y. Aharonov, S. Popescu, and J. Tollaksen,
	Phys. Today 
	{\bf 63}, 27 (2010). 






\bibitem{Nagao:2012mj} 
  K.~Nagao and H.~B.~Nielsen,
  Prog.\ Theor.\ Exp.\ Phys. {\bf 2013}, 023B04 (2013). 


\bibitem{Nagao:2012ye} 
  K.~Nagao and H.~B.~Nielsen, 
Proc. Bled 2012: What Comes Beyond the Standard Models, pp.~86-93 (2012) [arXiv:1211.7269 [quant-ph]].  






\bibitem{Nagao:2013eda} 
  K.~Nagao and H.~B.~Nielsen,
Prog.\ Theor.\ Exp.\ Phys. {\bf 2013}, 073A03 (2013). 




\bibitem{Nagao:2017book} 
  K.~Nagao and H.~B.~Nielsen, 
{\it Fundamentals of Quantum Complex Action Theory} (Lambert Academic
Publishing, Saarbr\"{u}cken, Germany, 2017).


\bibitem{Fukuma:2013mx}
M.~Fukuma, Y.~Sakatani and S.~Sugishita,
Phys. Rev. D {\bf 88}, 024041 (2013).






\bibitem{Mandela_effect} 
Ari Spool, The Mandela Effect, in KnowYour Meme, (Literally Media Ltd.), (available at:
http://knowyourmeme.com/memes/the-mandela-effect, date last accessed September 25, 2017). 




\bibitem{Nagao:2015bya} 
  K.~Nagao and H.~B.~Nielsen,
Prog.\ Theor.\ Exp.\ Phys. {\bf 2015}, 051B01 (2015).



\bibitem{Nagao:2017cpl} 
  K.~Nagao and H.~B.~Nielsen,
Prog.\ Theor.\ Exp.\ Phys. {\bf 2017}, 081B01 (2017).






\bibitem{Nagao:2017ztx} 
K.~Nagao and H.~B.~Nielsen, 
Reality from maximizing overlap in the future-included theories, 
to be published in Proc. Bled 2017: What Comes Beyond the Standard Models, 
arXiv:1710.02071 [quant-ph].





\end{thebibliography}
\end{document}